\documentclass[10pt]{article}
\usepackage[mathlines]{lineno}
\usepackage{amsmath}
\usepackage{amssymb}
\usepackage{graphicx}
\usepackage{cite}
\usepackage{color} 
\usepackage{setspace} 
\usepackage[labelfont=bf,labelsep=period,justification=raggedright]{caption}

\makeatletter
\renewcommand{\@biblabel}[1]{\quad#1.}
\makeatother

\date{}

\newcommand{\z}{Z\kern-0.6emZ}

\newcommand{\av}[1]{\left<{#1}\right>}
\newcommand{\bo}[1]{\boldsymbol{#1}}
\newcommand{\be}{\begin{equation}}
\newcommand{\ee}{\end{equation}}
\newcommand{\E}{I\kern-0.3emE}

\newcommand{\dd}{\mathrm d}

\newcommand{\td}[2]{\frac{\dd #1}{\dd #2}}
\newcommand{\nn}{\nonumber \\}
\newcommand{\half}{\frac{1}{2}}

\usepackage{bm}

\begin{document}

\begin{flushleft}
{\Large
\textbf{For whom will the Bayesian agents vote?}
}
\\
Nestor Caticha$^{1,\ast}$,
 Jonatas Cesar$^{1}$,
Renato Vicente$^{2}$, 
\\
\bf{1} Dep. de F{\'\i}sica Geral, Instituto de F{\'\i}sica, Universidade de S\~ao Paulo, CP 66318, 05315-970, S\~ao Paulo-SP, Brazil
\\
\bf{2} Dept. of Applied Mathematics, Institute of Mathematics and Statistics, 
University of S\~ao Paulo, 05508-090, S\~ao Paulo-SP, Brazil
\\
$\ast$ E-mail: nestor@if.usp.br
\end{flushleft}

\section*{Abstract} 
Within an agent-based  model where moral classifications are socially learned, we ask if a population of agents behaves in  a way that may be compared with  conservative or liberal positions in the real political spectrum. We assume that agents first  experience  a  formative period,  in which they adjust their learning style acting as  supervised Bayesian adaptive learners. The formative phase  is  followed by  a period of social influence by reinforcement learning. By comparing data generated by the agents with data from a sample of 15000 Moral Foundation questionnaires we found the following. 1. The number of information exchanges in the formative phase correlates positively with statistics identifying liberals in the social influence phase. This is consistent with recent evidence that connects  the dopamine receptor D4-7R gene, political orientation and early age social clique size. 2. The learning algorithms that result from the  formative phase vary in the way they treat novelty and corroborative information with more conservative-like agents treating it more equally than liberal-like agents. This is consistent with the correlation between  political affiliation and the Openness personality trait reported in the literature.  3. Under the increase of a model parameter interpreted as an external pressure, the statistics of liberal agents resemble more those of conservative agents,  consistent with reports on the consequences of external threats  on measures of conservatism. We also show that in the social influence phase liberal-like agents readapt much faster than conservative-like agents  when  subjected to changes on the relevant set of moral issues. This suggests a verifiable dynamical criterium for attaching liberal or conservative labels to groups.


\section{Introduction}
A central controversy in moral psychology 
and sociology deals with understanding the variety of moral values and whether  adherence to one set 
or another 
have  a genetic origin or arise from social interactions.
Political affiliation has
been associated to social interaction, to genetics 
 and to the combination of both 
(e.g. \cite{Hyman1959, Fowler2008, Alford2005,Hatemi2010a,Hatemi2011a}).  We address  questions about early age socialization, 
cognitive styles
and political orientation
within a  Moral Foundation theory (MFT) perspective using agent-based modelling and techniques from information theory. The present work is culturally situated within the fields of sociophysics \cite{galam,castellano} and computational social sciences \cite{epstein1,epstein2,gilbert} and is a companion to our previous work \cite{Vicente2009,Caticha2011,Vicente2012}.

In a series of papers  Haidt and coworkers 
\cite{Haidt2001a,Haidt2004a,Haidt2007a,Haidt2009a, Haidt2010a, Graham2009a, Graham2011a}
have described MFT,
an empirically driven theory dealing with the
foundations of moral psychology. It aims to understand
statistically significant differences in moral valuations 
of social issues and their
association to  coordinates of the political spectrum. 
The core tenet of the theory is that moral issues, which are valued 
mostly in an intuitive manner, 
can be parsed into a number of discrete dimensions, at least
five, possibly six or even more.  
According to Kohlberg \cite{Kohlberg71a,Kohlberg89a}  and 
Gilligan \cite{Gilligan82a}  dimensions representing care/harm
and fairness/cheating should be enough to span
the space of moral issues. Shweder {\it et al} 
\cite{Shweder1997} argued that
the dimensions should be three instead.  

The MFT states that  dimensions representing loyalty\-/betrayal, 
authority\-/subversion and sanctity\-/degradation should also be included in the moral space. The care\-/harm and fairness\-/cheating dimensions
are statistically more important for liberals than the rest, and each dimension of
the entire set is  of similar importance for conservatives.  Culture wars would be a consequence of these differences.

Consideration of other political cultures, such 
as libertarians leads to yet other dimensions, such as liberty/oppression
\cite{Iyer2012a}.
Political affiliation is also correlated with some characteristics of the Big Five
personality traits. Openness and liberal values appear together
frequently while Conscientiousness and conservatism are positively associated (e.g.\cite{Geber2009a}).
Further  associations between cognitive learning styles and political affiliation  have been suggested 
by  EEG  experiments \cite{Amodio2007}. 

In constructing the Motivated Social-Cognitive perspective Jost {\it et al.} 
\cite{Jost2003b,Jost2011a} make the 
assumption ``that conservative ideologies - 
like virtually all other belief systems -
are adopted in part because they satisfy some psychological needs''. We have also followed in our previous work
 \cite{Caticha2011, Vicente2012}  a motivation driven approach
with a totally different methodology: studying mathematically the dynamics of 
agent-based models using information theory. We considered the discomfort 
associated to disagreement \cite{Eisenberger2003} and the motivating
pressure was to reduce pain associated to social exclusion. 
This was implemented by a  learning dynamics designed to maximize a utility function or, equivalently,   minimize an energy-like function. 
Haslam \cite{Haslam} correctly argues that
not all social figuring is or should be a matter of 
cost/benefit calculation. 
In a third person description, within a mathematical language,
we calculate, but the social agent does not calculate, it just acts. 

In our previous approach we characterized in a simplified society of agents the effects of different learning styles on the statistics of their opinions about  a set of issues. We will call the {\it artificial data set} to the data obtained by simulation of the agents.  
The analytical and numerical results were compared 
to  data gathered by the Moral Foundation 
Questionnaire
project of Haidt and collaborators \cite{HaidtData},  to which we will
refer as the {\it empirical data set}.  Agents learning with 
an algorithm that treated new and corroborative information in the
same way, exhibited (a) less dispersion of opinions, (b) longer times to 
readapt under changes of the issues under discussion and (c)  histograms
of opinions very similar to those of self declared conservatives in the
empirical data set. On the other end of the spectrum of cognitive styles, agents that
could be thought to score higher in an Openness personality trait, 
since they gave more importance to new data than to corroborative data, 
(a) showed greater  dispersion of opinions, (b) readapted faster after changes of the issues  and (c) were statistically similar to 
self declared liberals.

Note that we avoided the difficult task of theoretically 
predefining conservative or liberal. We just took a pragmatic route,
comparing the results of our model with empirical data where subjects 
had declared their belief about their  positions in the political spectrum.
In other words, a society of agents is classified as  conservative or liberal
by the proximity of their statistical 
signatures to those obtained from  the Moral Foundation Questionnaires  of groups who believe and declared to be of a certain political affiliation.

In this paper we address the following question: why are different 
cognitive strategies present
in the population? Distal causes could be such as the advantages
of societies with a higher
cohesive set of values due to conservatives and shorter
readaptations due to liberals. If we ask for more proximate causes, genetics or 
heterogeneous social interactions are possible explanations.
 A discussion by Smith et al \cite{Smith2011a}
illustrates the long path between genetics and opinions
about specific issues, including four intermediate levels: biological,
cognitive/information processing, personality/values and ideology with
the environment influencing each one.

Fowler and collaborators presented  evidence for interactions between 
genetics  and politics. In  \cite{Dawes2009a} they
link the DRD2 dopamine receptor to partisanship hereditability.
More relevant to our present study, is their analysis of data 
\cite{Settle2010b} 
from the National Longitudinal Adolescent Health
study indicating that a certain 
allele (7 repetitions long allele) of the dopamine receptor gene 
DRD4 may have just that kind of influence. For those having
two copies of the allele, the number of friends during early age
condition the probability of their self declared political 
affiliation as an adult. The direction is such that those that
had a larger number of 
friends are associated to a larger  probability of being a liberal
as an adult. 

Here we aim at explaining the diversity in moral valuation within our agent based framework
 by adopting an information theory point of view, in particular we 
consider an  artificial society composed by interacting Bayesian information
processing agents. Each 
agent has a set of social neighbors and exchanges information 
in the form of opinions about issues. 
Learning means that
when the information brought by the  opinion of a social neighbor arrives, 
there are certain changes in the weights attributed to each moral dimension.

The main results about the learning process following from this approach are two. First,
that the learning algorithm is not static but adaptive. It
depends on the number of opinions to which  
an agent has been exposed in social interactions. Second, that 
for different numbers of such opinions,   
the difference of the ensuing learning algorithms
can be described by the different modulation given to opinions that carry
novelty of information relative to opinions 
that carry corroborative information.
Figure \ref{fmod}-left shows  the modulation function for different number of social interactions. The modulation function is  a measure of the 
overall scale of the changes of the weights for the moral dimensions elicited by a 
particular issue and opinion of a social partner. Including
 the possibility of errors in communication, 
a Bayesian learner with the information that there might be imperfect
communication,
acts with suspicion and ignores disagreeing opinions on issues about which
it has a strong opinion (Figure \ref{fmod}-right).

The agents of our model are Bayesian during an early window of time we call the {\it formative phase}. 
Each `young' agent is  exposed to a random number of social information exchanges.
At the end of the formative phase the learning algorithm stops evolving and agents enter the {\it social influence phase}.
Agents, each with its particular 
fixed learning algorithm determined by the random socialization in the formative phase, 
exchange information about a set of issues and continue learning. After a time where a steady
state has been achieved, we collect statistical information about the state 
of the society in the form
of histograms of opinions (the artificial data set, ADS). A similar set of statistics can be extracted from 
the set of questions about moral issues (the empirical data set, EDS) collected by Haidt and
collaborators from the Moral Foundation project \cite{HaidtData} as done
in  \cite{Caticha2011, Vicente2012}.
Numerical comparisons of the statistics
 permit identifying a class of agents with a group of respondents
with  a given declared political affiliation. The conclusion is that
the number 
of opinion exchanges in the formative phase is correlated with the political affiliation of the
corresponding group of the responders. Agents with large number 
of opinion exchanges in the formative phase are identified with liberals after the social influence phase, 
those with a small number are identified with conservatives.

In section \ref{methods} and  appendices we present the mathematical aspects of the
theory, first the Bayesian algorithm of learning that evolves during 
the formative phase, then the description of the social influence phase where agents interact.
 The rest 
of the paper has
a descriptive approach where no mathematical formalism is used. 
In section \ref{results} we present the results and describe
the comparison to the data obtained from the Moral Foundation 
questionnaires. We end this paper with a discussion of the 
results, the limitations of the theory and possible extensions.

\section{Methods \label{methods}}
\subsection{Formative phase}
Here we describe within a Bayesian framework 
the way agents process information. We suppose  that
issues are parsed into a set of five numbers. An issue labeled $\mu$ is
represented by
$\bo{x}_\mu=\left\{x_{a\mu}\right\}_{a=1,...5}$, each $x_{a\mu}$ describing the
bearing of its content on a moral dimension. Agents emit opinions
in a fast, automatic, intuitive  manner independently 
of intricate if-then rules. In the model this is 
done by summing over the five dimensions
the content of each moral dimension of the issue, weighted
by  the importance the agent attributes to each foundation.
 The moral state of agent  $i$ at time $t$,
called the moral matrix in MFT, is also a vector 
$\bo{\omega_i}(t)=\left\{\omega_{a,i}(t)\right\}_{a=1,...5}$.
The opinion of agent $i$ about issue $\mu$ is
$h_{i,\mu}=\sum_{a=1}^{5}{\omega}_{a,i}{x}_{a,\mu}$ 
and its sign $\sigma_{i,\mu}=
\mathrm{sign}(h_{i,\mu})$ shows whether an agent is for or against 
an issue.

During a social encounter in the formative phase
an agent $i$ receives information $y_\mu=(\sigma_{j,\mu},\bo{x_\mu})$  
emitted by the social partner $j$. Learning occurs in order to 
decrease disagreement over issues. Within this learning scenario,
we hypothesize that evolutionary pressures to increase the prediction
of the opinions of others would select learning algorithms near Bayesian optimality (see \cite{Neirotti2003}).  
As shown in Appendix \ref{model} the resulting learning 
algorithm that approximates a full Bayesian use of the
available information, can be described in two different ways. One as
 a motivational 
algorithm where a cost or energy like function ${\cal E}$ is decreased by the
changes elicited by learning.
The other 
as a modulated Hebbian learning algorithm with the central 
concept, the modulation function $ F_{mod}$ (figure \ref{fmod}), being a measure of the 
importance attributed to a given issue and the opinion
of the interlocutor. In terms of the moral matrix 
$\hat{\omega}_{a}(t)$ and a measure of the full social experience 
$C(t)$, both ways are:
\begin{eqnarray}
\hat{\omega}_{a,i}(t+1) &=& \hat{\omega}_{a,i}(t) - x_{a,\mu+1} \sigma_{j,\mu+1}
C(t) \frac{\partial{\cal E}_\mu}{\partial z_\mu},\label{dinamicaE} \nn 
& =& \hat{\omega}_{a,i}(t) +x_{a,\mu+1} \sigma_{j,\mu+1}F_{mod}, \\ 
C(t+1) &=& C(t)- C(t)^{2} \frac{\partial^2{\cal E}_\mu}
{\partial z_\mu^2},\label{dinamicaF}\nn
&=& C(t)+C(t) \frac{\partial F_{mod}}
{\partial z_\mu}.
\end{eqnarray}
The modulation function and the cost are related by
$F_{mod}(z,C)=-C(t) \frac{\partial{\cal E}_\mu}{\partial z} $ 
where $z_\mu=\sigma_{j,\mu+1} h_{i,\mu}$ measures the concurrence/disagreement 
between agents $i$, the receiving agent, and agent $j$
the opinion emitting agent. $C(t)$ is related to the width of the posterior
distribution and decreases as learning occurs. 
We also use $\rho(t)= 1/\sqrt{1+C(t)^2}$, a convenient variable since it
takes values between zero and one.  It is close to zero 
when an agent had a small number of social encounters and 
approaches one as the number increases. Hence the modulation function and the cost are functions of $z$ and $\rho$.

The main results of this paper derive from 
the fact that the modulation function of
the Bayesian algorithm (1) is not the same throughout the learning
period and changes as more information is incorporated and
depends on the number of social encounters; and (2) it depends 
on the novelty that the opinion of the social partner carries. These two aspects are clear in figure \ref{fmod}.Right, 
where the modulation function is plotted as a function of $z=h_i \sigma_j$, for different fixed values of $\rho$, which measures the number of social interactions. Note that $z= |h_i|\sigma_i \sigma_j$ measures the strength $|h_i|$ of the  opinion held by $i$ and the  $\sigma_i\sigma_j$ which is positive if the opinion $\sigma_i$  prior to learning  agent $i$  is the same as the that of agent $j$ and the information is corroborative, and $z<0$ if the opinions are opposite and
the arriving information is considered a novelty.

\subsection{Social influence phase }
We consider  the number of information exchanges or socialization events
in the formative phase as a random number,  not the same for all agents  
and thus the effective $\rho$ for
each agent is a number between zero and one.  The agents in the formative phase
learned to learn and now they just learn from each other with a frozen modulation
function. The validity
of this  supposition as something that
represents the development of adolescents  has to be investigated
in an independent way. 
It loosely rings with Piagetian overtones \cite{Piaget65}.
We also consider the fact that people tend to interact with
the likes \cite{Abrams1990}. So we consider as a nonessential simplification, a system of agents all 
with the same $\rho$ each one in a site of a social lattice, exchanging
information
and then investigate the effect of changing $\rho$.
The dynamics of information exchange is analogous to that considered in 
\cite{Vicente2009,Caticha2011, Vicente2012}, the only difference being that 
the learning occurs with the Bayesian algorithm described above. 

We suppose that a society discusses a set of $P$ issues. 
Parsing of an issue into a vector might be {\it subjective},
expressed by the fact that agent $i$ obtains a vector $\bm x_i$. 
Exchange of information between 
agents is about the average vector
\begin{linenomath}
\begin{equation}
\z \propto \frac{1}{P}\sum_{\mu=1}^P \bm x_i^\mu,
\end{equation}
\end{linenomath}
which we suppose reasonable to be independent of the agent, since
fluctuations due to subjective parsing, if unbiased, tend to cancel out. 
We call $\z$ the Zeitgeist vector since it captures the contributions
of all issues that are currently being discussed by the model
society. Without any loss
it will be normalized to unit length. 
The opinion of agent $k$ about the
Zeitgeist is 
\begin{linenomath}
\begin{equation}
h_k= \z \cdot \bm w_k
\end{equation}
\end{linenomath}
and its  sign  is denoted by $\sigma_k=\mathrm{sign}(h_k)$. 
We now consider a Metropolis-like stochastic dynamics of information exchange.
The 
conjugate parameter $\beta$, determines the scale of tolerance
to fluctuations in the cost ${\cal E}$, that is, it determines
how important it is to conform to the opinions of others agents and eventually
sets the scale of fluctuations
of an agent's moral vector around the Zeitgeist.

\subsection{Simulation}
The artificial data is generated by the following procedure. We suppose that agents are characterized by a learning algorithm parametrized by $\rho$ depending on the number of social interactions they experienced during the formative phase (see appendices A and B for details).  We also suppose that 
agents only interact with counterparts holding equal $\rho$. We choose a random social undirected graph from an ensemble here taken to be generated by a Barabasi-Albert model with $N=400$ and $m=10$. Our results are not strongly dependent on the details of the social graph topology \cite{Vicente2012}.

Agents start the social influence phase with moral weights that are represented by unitary vectors $\bo{\omega_i}$ with  random positive overlaps with a fixed Zeitgeist vector $\z$. The social influence dynamics is implemented as a Markov Chain Monte Carlo process as follows. At each step an edge $\langle i j\rangle$ of the social graph is randomly and uniformly chosen. One of its vertices (let it be $i$) is then  marked as the influenced agent with probability $1/2$. The influenced agent chooses  a random unit vector  $\bo{\omega^\prime_i}$ and changes her moral weights  $\bo{\omega_i}$  with probability given by $\mbox{min}\{1,\exp\left(-\beta\Delta{\cal E}\right)\}$, where $\Delta{\cal E}=\sum_{ j \in {\mbox{\tiny neigh}}(i)} \left[ {\cal E}(h_i,\sigma_j) - {\cal E}(h^\prime_i,\sigma_j) \right]$. Note that the agent has complete  access to his opinion $h_i$, but only knows the sign of the influencer opinion $\sigma_j$. Observe also that the pressure parameter $\beta$ regulates the acceptance rate in the transition. High pressure $\beta$ makes moral representation changes more difficult. 

\begin{figure}
\begin{center}
\includegraphics[scale=0.6]{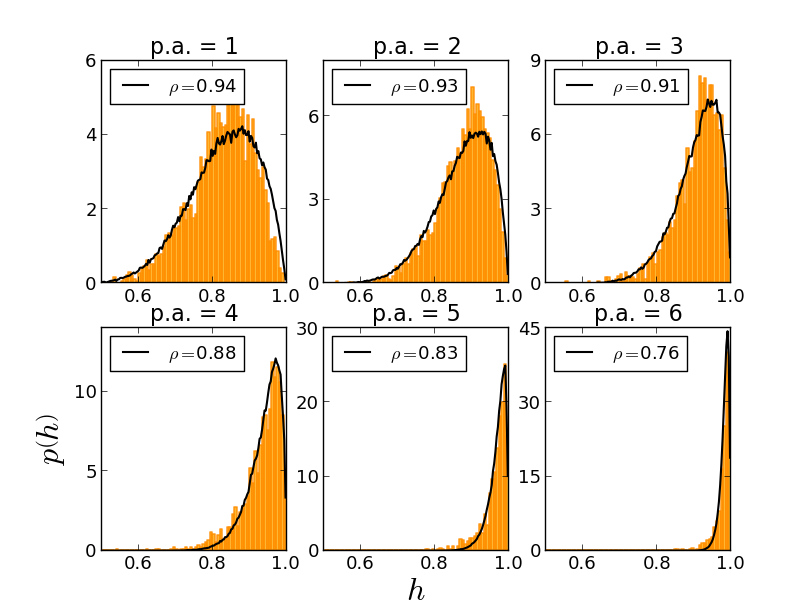}
\end{center}
\caption{{\bf Comparison with empirical data}. Empirical opinions (histograms in orange) correspond to the overlap between moral weights obtained by MFT questionnaires and the average weight of the most conservative group (Zeitgeist direction $\z$). The histogram obtained by simulating social influence in a social network with homogeneous learning styles (homogeneous $\rho$) and computing  overlaps between  moral weights of a the agents and a given Zeitgeist direction ($\z=(1,1,1,1,1)$ in the simulation) are represented by the black line. In each graph we find $rho$ that best fits the empirical histogram for pressure $\beta=3.8$ and for each political affiliation group. Simulations are performed on a Barabási-Albert network with $N=400$ and average degree $20$.}
\label{histograms}
\end{figure}

Data are collected after the system reaches  equilibrium. We typically wait $T_{\mbox{\tiny term}}=6\times 10^4 N$ interactions before gathering uncorrelated   samples for time averaged opinions $h_i$ that are used to build  the histograms depicted in Figure \ref{histograms}. To guarantee that samples are uncorrelated we calculate autocorrelation times $\tau$ and then select properly spaced $T_{\mbox{\tiny term}}/\tau$ samples. The whole procedure is repeated a $n$ times until $500$ independent samples are drawn ($n=4$ being the minimum for the data we report). Our codes and preprocessed data at available at \cite{githubJonatas}. Raw data for the  Moral Foundations survey can be obtained from \cite{HaidtData}.

\subsection{ Confrontation between artificial and empirical data}
A society of agents is characterized by the values of $\rho$, measuring the 
effective socialization in the formative phase, and of $\beta$ that sets the
pressure on the society during the social influence phase. While in a society different agents with
different $\rho$'s and feeling different $\beta$'s will interact, it is 
a reasonable first approximation to consider that people will more likely
interact in a meaningful manner with those that are more similar. 

In a steady state of a society of agents, changes in the moral matrices still occur, 
but the distribution $P_{ADS}(h|\rho, \beta)$ of opinions about the Zeitgeist are stable in time. 
From $15000$ MFT questionnaires (see \cite{Vicente2012} for a description of this data set)  we obtained the 
data \cite{HaidtData} and the following information.
(i) $\{w_a\}_{a=1...5}$, 
the (normalized) weights of the moral matrix and the political affiliation of 
each respondent. (ii) The empirical Zeitgeist vector ($\z_e=\{Z_a\}_{a=1...5}$) defined as the
average weight vector of the most conservative group. 
(iii)  The empirical Zeitgeist opinion $h^e=\sum_a w_a Z_a$ for each respondent.
(iv) The empirical distribution of opinions $P_{EDS}(h|pa)$ is obtained for each 
of the political affiliations $pa$.

A distance between the two distributions is measured  
\begin{linenomath}
\begin{equation}
D(\rho, \beta; pa)=\sum_{h \in bins}\left(P_{ADS}(h|\rho, \beta)-P_{EDS}(h|pa)\right)^2.
\end{equation}
\end{linenomath}
by summing the quadratic difference over a set of bins of $h$.
Figures  \ref{histograms} and \ref{politicalaffiliation}  are obtained
by identifying the value of $pa$ for the regions on the $\rho-\beta$ space where
$D(\rho, \beta; pa)$ is smallest. If the smallest $D(\rho, \beta; pa)$
is larger than a threshold value of identification (e.g 0.1) then the point is not identified
to any political affiliation.

\section{Results \label{results}}

\subsection{Learning dynamics}
We started with Bayesian learning and obtained two equivalent
descriptions of the learning  {\it dynamics } describing changes in the weights
of the moral dimensions.  
The dynamics described in equation \ref{dinamicaE}
 can be seen to be a gradient descent: changes of the weights
are in the direction of decreasing  a quantity ${\cal E}$
that can be interpreted as an energy, a cost or a pain. 

We claim that this  motivational (or utilitarian) form of learning can be useful to understand better what is occurring. 
Then for each example the change occurs in the direction
which tends to reduce the  error of classification, to 
increase conformism or to reduce pain derived from disagreement.  
But it is just a mathematical fact that may go along uninterpreted
and be described just as a Bayesian inspired learning. We can describe the
falling rock as moving along a trajectory that decreases potential energy.
It is not the rock that is being utilitarian or motivated to reduce an energy, but it is our description
using energy that seems utilitarian. The motivation lies in our third
person description. 

\subsection{The modulation function}
By using the idea of the
modulation function we described (eq.\,\ref{dinamicaF})
the same learning dynamics differently.
The modulation function  measures the importance
of the information carried by the example. It could be thought
in a loose way as representing the signal from  something like an 
amygdala, which would signal more strongly in case the
example causes surprise due to the novelty of an unexpected result.

In addition to measuring surprise, it is striking that
it depends on $\rho(t)= 1/\sqrt{1+C(t)^2}$.
What is striking about a $\rho$ dependent 
modulation is that in a static scenario and for an agent with only
one social partner we can prove \cite{Kinouchi92} that $\rho$ increases with the
number of information exchanges, and this  still holds numerically
when learning from several correlated social partners.

We now analyze the case shown in Figure \ref{fmod}-Left
for noiseless communication. 
 At the beginning of the learning
process the modulation function is flat. Every piece 
of information, every example
receives the same modulation. Being right or wrong is of little consequence
in the manner in which the information is incorporated.
As learning occurs, from the
information exchange with social partners,  the modulation
function decreases for positive $z$ and increases for negative $z$.
Examples that carry new information start getting a higher modulation.
Those that were predicted correctly, are less effective in 
fostering changes in the weights of the moral vector.
Examples carrying new information make larger impacts, 
those that corroborate the opinion of the agent, have a smaller 
influence. 
As $\rho$ increases this effect is amplified. 

In Figure \ref{fmod}-Right a noisy communication channel 
is introduced. With probability $\epsilon$ (equal to $0.2$ in the figure),
the received  opinion is flipped. But the agent doesn't know
which specific examples are corrupted. The Bayesian algorithms permits 
incorporation of this information, the result is a distrust 
effect. If the agent is very sure about its opinion (large absolute value 
of $z$), but
it differs from that of the social partner ($z<0$), it
tends to disregard the example by doing smaller changes in the
weights. This increases with the value of the noise level and with
$\rho$.

To sum it up, the modulation has three characteristics which
we list in decreasing order of importance. The modulation function 
depends on
\begin{enumerate}
\item  Novelty/Corroboration: a measure of whether the  example 
carries new information ( $z <0$) or is corroborative ($z>0$),
\item Socialization in the formative phase: a measure of the number of information exchanges ($\rho$),
\item Trust/distrust: a measure of the reliability attributed to the social 
partners.  Given $\epsilon$, if $z$ is too negative, the example is 
not considered new information but rather it is distrusted and its
effect is small.
\end{enumerate}

We have analyzed the simple dynamics where the covariance is represented
by a single parameter $C$ or equivalently $\rho$ (see appendices for details). This is probably a
good approximation but it is reasonable to assume that the dimensions
may be interdependent. For example caring for a member of the group
may be larger than for a member of another group, also cheating an
authority figure may be different than cheating a common member of the group. 
This can be modeled by off diagonal terms in the covariance matrix, that mix the moral 
dimensions of care and loyalty, in the first case or the fairness and authority
dimensions in the second. It is not clear at this point
whether this means that there are different neural circuits that deal 
with the dimensions but are interconnected, or if there exists
combinations of dimensions that are independent. This last option 
seems attractive from a mathematical point of view, being just another 
case where diagonalization is useful. However the existence of interacting
circuits is probably more in accord with the fact that evolved
language attributes specific names to them and not to their combinations. 

\subsection{The political affiliation of Bayesian agents}

Agents, of course, do not have a political affiliation. 
However we can measure the distribution of opinions $P_{ADS}(h_i|\rho,\beta)$ after the formative phase, about the 
Zeitgeist vector for a society of agents all with the same value
of $\rho$ and pressure $\beta$. Now we  have a statistical signature that can be
compared to a similar signature extracted from the data of the Moral
Foundation questionnaires for each political affiliation group. This is similar to the methodology
we  used in \cite{Caticha2011} and \cite{Vicente2012}. 
This results in the identification for fixed $\beta$, 
of the measure $\rho$ of the formative phase, and the
self-declared political affiliation of the respondents of the questionnaires.
This is done for several values of $\beta$ and the result
is shown in Figures  \ref{histograms} and \ref{politicalaffiliation}. It is clear that 
the populations of agents with small value of $\rho$, or small number of social 
information exchanges, are close to conservatives and those populations
with high $\rho$ or large number of social information exchanges, are more
likely to be identified with liberals. Note that this is not a
one to one identification. We are not saying that a given agent's
value of $\rho$ determines political affiliations, but rather that
this subset of the population will  have  a distribution of opinions consistent
with such identification.

\subsection{The phase diagram}
\begin{figure}[ht]
\begin{minipage}[b]{0.5\linewidth}
\centering
\includegraphics[width=\textwidth]{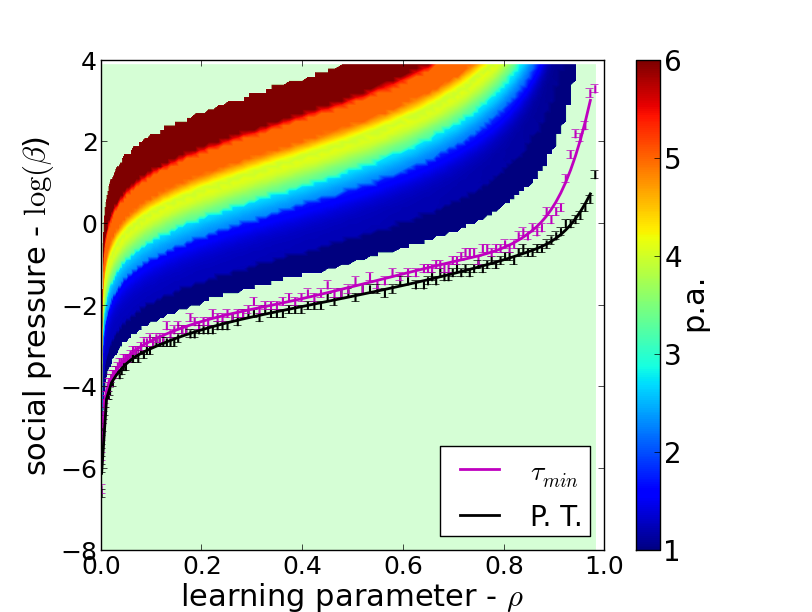}
\end{minipage}
\hspace{0.3cm}
\begin{minipage}[b]{0.5\linewidth}
\centering
\includegraphics[width=\textwidth]{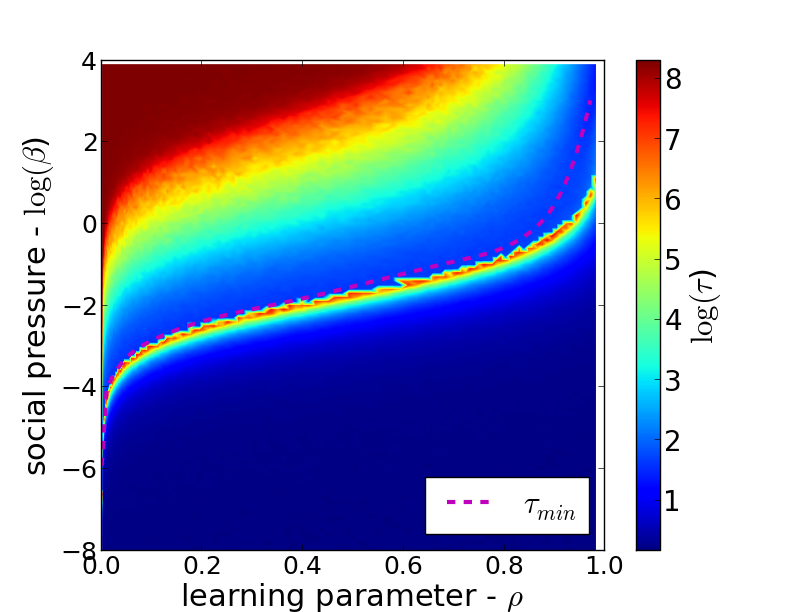}
\end{minipage}
\caption{{\bf Phase diagram and relaxation times.} Left: The phase diagram in the space $\rho$, a measure of 
the complexity of socializations versus $\beta$ the 
pressure. The
stripes represent regions of the space of parameters where agents 
could be statistically identified with a group with a given political 
affiliation. The lower line represents the boundary between order (above)
and disordered (below) societies. Below the transition line, and for very large $\beta$, 
no identification with MFT questionnaire respondents was found. 
Right: Color coded relaxation times after 
changes in the set of moral issues. Note that at the transition
relaxation times are very large. This is called critical slowing down. 
For the agents identified with respondents of the MFTQ, the lowest times
correspond to those liberal identified agents and the largest times 
to conservative identified agents. The line just above the transition
shows the locus of minimum correlation time as a function of $\beta$, 
for fixed $\rho$.  }
\label{phasetempo}
\end{figure}

The phase diagram is the instrument used to represent the variety
of possible  collective behaviors of systems composed of many interacting
units, in particular a society of agents. 
The phase boundary separates regions of totally different
properties. 
In Figure \ref{phasetempo}-Left we show the phase diagram in
the space of parameters ($\rho,\beta$). Above the transition 
line which is the lower line,  the society of
agents has an ordered phase. That there is some coherence of opinions
in the society is shown by the fact that the average value
of the opinions is not zero. Below the transition  line the model society
is disordered in the sense that opinions are so varied that
they average to zero, opinions are not shared, the Zeitgeist is not clear.
This phase, the opposite of the ordered phase, resembles the anomie
of Durkheim. 
The stripes follow regions where the statistical signatures
are similar to  a certain political affiliation group.

\subsection{Readaptation times} 

What is it that conservatives conserve?  
If a society of agents identified with conservatives (low $\rho$) were
to readapt after changes faster than one identified with liberals, our theory would 
have to be thrown away. But it is a result of our theory that
liberal-like societies are faster than conservative-likes in readapting. 

Several approximately equivalent ways of defining relevant measures
of readaptations times can be introduced and we have looked at two
such measurements and obtained similar results.
After a steady state was achieved
and the steady state distribution $P_{ADS}(h|\rho, \beta)$ is measured, 
the Zeitgeist $\z_{old}$ is changed to a new Zeitgeist $\z_{new}$.
Call this time $t=0$. 
After a sweep of information exchanges of all the agents, $t$ increases
by one unit,
the distribution of opinions about the new Zeitgeist $P_t(h)$ is 
measured. A distance between the two distributions is measured  
\begin{linenomath}
\begin{equation}
D(t)=\sum_{h \in bins}\left(P_t(h)-P_{ADS}(h|\rho,\beta) \right)^2.
\end{equation}
\end{linenomath}
by summing the quadratic difference over a set of bins.
As usual the relaxation is exponential so we parametrize  $D(t)=D_0 e^{-t/T}$
in terms of the adaptation time $T$ which depends on $\rho$ and $\beta$.
For more about this measure see \cite{Vicente2012}.

A second possibility is the correlation time, defined by measuring the
decay in time of the time correlations. These are defined by
the difference between the expected value of the product of the 
moral vector at two different times and the product of their 
expected values  
$c(t)=<m(t'+t)m(t')>-<m(t'+t)><m(t')>$
which also decays exponentially as $c(t) \propto \exp (-t/\tau)$
with a characteristic relaxation time $\tau(\beta,\rho)$.
Figure \ref{phasetempo}-Right
shows the result of measuring numerically the relaxation times of
the different populations.
At the transition line the relaxation time
grows beyond any limit as ever increasing
populations are considered. This is called critical slowing down.
Bellow the transition, the system is disordered and after the
Zeitgeist change, it returns very quickly to the steady state. 
This temporally and spatially uncorrelated
region is uninteresting from the point of view that 
no data for an analogous population  in this region is currently available. 
Above the transition line, as we move to larger $\beta$ with 
fixed $\rho$, $\tau$ goes down very fast, attaining its
minimum values near the regions of ultraliberals and then
has a monotonous rise into the conservative region. 
The shape of the regions where $\tau$ remains approximately constant
are similar in shape to the regions in the phase diagram identified with
a given political affiliation. This suggests that  political affiliation 
could be empirically characterized by collective 
relaxation time that  increases as the political spectrum is transversed 
from groups conventionally labeled as liberals 
to those labeled as conservatives. 

\subsection{Threats: Conservative shift under increase of pressure}
The pressure parameter $\beta$ determines 
how important it is to conform to the opinions of others. A more 
detailed modelling of the agents  could  make a difference
between informational or normative peer pressure. Or the differences between 
situational or dispositional attributions of $\beta$.
Economical or environmental pressures could
influence how the the social environment is perceived. 
Coarsely, $\beta$ describes the overall motivation that
sets the scales of adaptation of beliefs. 
However it is set, it controls how 
strongly  the agent should conform to other
opinions or to the overall current Zeitgeist. 
Technically, it determines the scale of tolerance
to fluctuations in the cost function ${\cal E}$.
Equivalently  $\beta$ sets the scale of fluctuations
of an agent's moral vector around the Zeitgeist.

\begin{figure}
\begin{minipage}[b]{0.5\linewidth}
\centering
\includegraphics[scale=0.4]{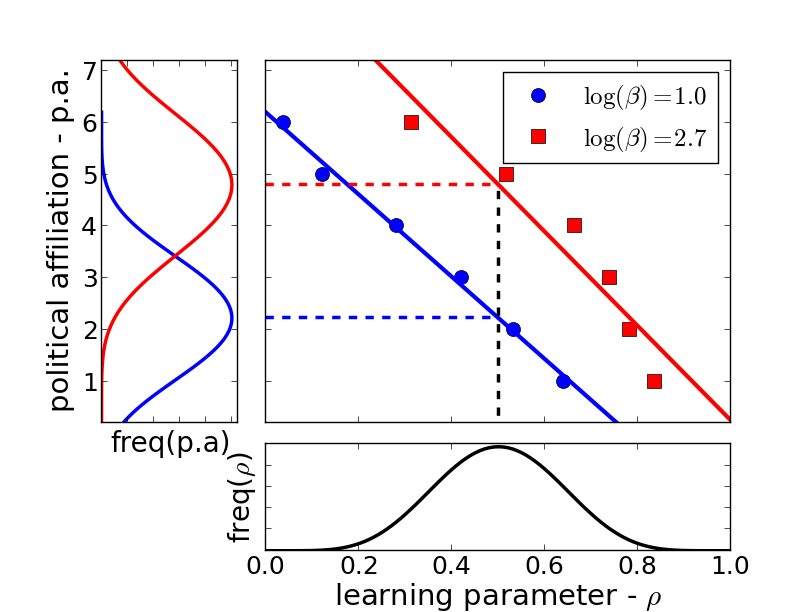}
\end{minipage}
\hspace{0.3cm}
\begin{minipage}[b]{0.5\linewidth}
\centering
\includegraphics[scale=0.4]{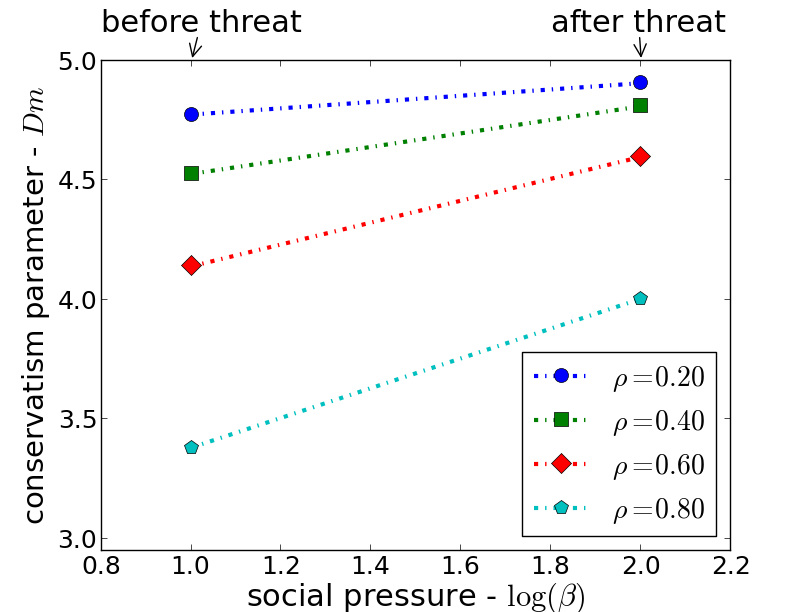}
\end{minipage}
\caption{{\bf Threats.}
Left: The number of effectively conservative agents changes with
pressure. If the population has a  distribution 
of social encounters as shown in the bottom panel, 
the resulting distributions of political 
affiliations changes, for different 
pressures 
as shown in the left panel.
Right: The effective number of  moral dimensions for two values of the
pressure, before and after an external threat. 
If a threat leads to increased 
pressure, the
statistical signature of liberals agents will
look more like that of conservatives. }
\label{distpa}
\end{figure}

We can model the 
effect of an external event that threatens the group to  which 
the agent belongs by considering that the pressure $\beta$ 
increases. 
The effects of the
threat in the political affiliation of the agents, shown in Figure
\ref{distpa}-Left is that the population will shift towards the
conservative end of the spectrum. 
We supposed a fixed distribution of  the number of social 
information exchanges $\rho$, and the effect on the distribution of 
political affiliations before and after a threat which increases the
peer pressure $\beta$. 
Our model predicts also that under
the perceived decrease of an external threat the populations
will shift towards the more liberal region.  

We have defined the effective number 
of moral dimensions of a group with a given political 
affiliation. This is done by averaging the weights over all 
members of the population and multiplying by  the number of moral
dimensions $d_m=5$. 
For groups of agents that are identified with conservatives,
the effective moral dimension is near 4.8 . For those identified with liberals
it is near 3.5. Both increase under increase of the peer pressure parameter
$\beta$ as shown in Figure \ref{distpa}-Right. This is in qualitative agreement with 
experiments reported in \cite{Bonanno2006a,Nail2009b,Nail2009c} and further
work in \cite{vandertoorn}.

\section{Discussion and Conclusions} 
The main characteristic of Entropic inference and the 
Bayesian approach to information theory is
that the mathematical structure to represent 
beliefs in the absence of complete information \cite{Jaynes2003, Ariel2008},
if manifest inconsistency is to be avoided, is probability theory. 
As presented in section \ref{methods} and in appendices \ref{model} and B 
a Bayesian study of the learning dynamics of moral classifications 
can be described as the changes in the weight parameters for
each dimension that lead to a decrease in the cost, interpreted
as psychological discomfort, caused by differences of opinions. 

The main result we presented here is that the cognitive style
of the Bayesian agent depends on the complexity of the social 
interactions in the formative phase and cognitive style induces a statistical 
association to political affiliation. The formative phase is a mimic of 
the pre-adolescent phase in the life of an agent and the 
social influence phase is a mimic of the post-adolescence. 
During the  social influence phase the
agent's cognitive style is crystallized, so that it 
 ceases to change, although the agent is still capable of learning, then
it follows that statistically the agents when identified with
respondents of the MFTQ, with the social complexity of the formative phase
being positively correlated with liberalism. This is exciting
since in  Settle {\it et al.}
 \cite{Settle2010b} the number of childhood friends is 
positively correlated to liberalism, at least for those that 
have two alleles of the DRD4-R7 gene. They cautiously withhold 
from claiming that a gene for political ideology was identified
and just claim that evidence points to a gene-environment interaction.

Within the context of \cite{Settle2010b} 
what is the genetic interpretation of our results? Our methods 
do not address this problem. Genetically, having two long R7 
DRD4 alleles, may contribute to making the number of friends a proxy for
social complexity in the formative phase. 
But some other genes may contribute to Openness, 
with influence on the number of friends, thereby influencing 
the cognitive style with respect to the differences of learning
novelty and corroborations.
But our approach does not address this mechanism
nor those by which other phenotypes become conservative
or liberal. What we say
is that Bayesian optimal learning predicts that
 number of social interactions in the formative phase 
will correlate with liberalism in the social influence phase. But, 
why should agents
be Bayesian optimal? An answer can be given 
based on the results of, first,  \cite{Kinouchi92} where
the functional optimization of the learning process was obtained, second
 \cite{Opper96}, where a related algorithm was shown to be  the
online version of the Bayes algorithm and third
 \cite{Neirotti2003} where, using  evolutionary programming, 
the authors showed that perceptrons
evolving under pressure for having larger generalization ability,
were driven to learning algorithms that resembled Bayesian optimal
algorithms both in functional form and performance. Thus 
if learning algorithms for moral classification from examples,
are subject to evolutionary 
pressures for better generalization, Bayesian optimal
like algorithms will be approximated. Cognitive styles will then depend
on social interactions at the earlier phase in the life of the agent. 
A question that remains is why there would be a 
formative phase, for learning how
to learn and setting overall parameters, 
and a social influence phase, for actual learning. These questions
are outside of our scope and will need other methods and inputs.

What are the predictions of the model? These are summarized in 
table \ref{table:predictions}. Relaxation times were
never used in the theoretical formulation of the problem. They
are  a physical consequence of the social information exchanges
and hence a prediction of the model. Different cognitive styles, through social 
interactions, lead to different adaptation times. 
The existence of a phase transition between an ordered moral phase 
and a moral disordered phase might not be observable since societies
morally disordered might not exist. However this model can 
be applied to other culturally relevant landscapes, where
groups on both sides of the divide might be found. A question that remains
is if in those contexts, pressure will lead to Bayesian optimality
resulting in cognitive style diversity. 

Another prediction of the model is that under an increase of $\beta$, the peer
pressure, a society as a group will tend to seem
statistically  more conservative,  as shown in Figure \ref{distpa}. 
This effect of peer pressure increase
might be behind the results of Bonanno and Jost
\cite{Bonanno2006a} and Nail {\it et al} \cite{Nail2009b}
about the increased conservatism of subjects that were exposed to the 9/11
attack. However  Nail {\it et al} \cite{Nail2009c}  show 
that there is no need for social interaction in order to become more 
conservative, suggesting that our interpretation of $\beta$ as peer 
pressure could be extended to a self-regulated parameter that is adjusted
dynamically from information about social context.

An empirical definition and consequent  measure of
pressure might be done following the methodology of
\cite{Gelfand2011a} where nations were classified on a tight/loose 
scale. Analysis of morality data sets for individual 
countries could point out if our  pressure and their
tight/loose scale are related. Since we use only USA citizens 
questionnaires, we are not able to address this question here, leaving
the issue for a forthcoming paper.

An important characteristic of our model is that it is semantically free.
Just or loyal in the mathematical space where the agents are
defined are concepts devoid of meaning. We believe that this aspect has to be addressed from
an evolutionary perspective in order to understand
the emergence of the dimensions
and hence provide our mathematical backbone of a semantic dressing.
\begin{table}[!ht]
\caption{{\it Predictions of the model of Bayesian Agents .}
Left: Variables refer to the Bayesian Agent Model. 
Right: Compared to the data from the Moral Foundations project.} 
\centering 
\begin{tabular}{||l|l||}
\hline
Variable         &  Prediction \\ \hline
Cognitive style: novelty $\approx$ corroborations & Correlates with conservatism \\ \hline
Cognitive style: novelty $>$ corroborations & Correlates with liberalism \\ \hline
Social complexity in formative phase $\rho$   &  Correlates with liberalism\\ \hline
Correlation times $\tau$ & Correlates with conservatism\\ \hline
Increase in pressure $\beta$ & Liberals look statistically more like conservatives
\\\hline
\end{tabular}
\label{table:predictions} 
\end{table}

\section*{Acknowledgments}
It is a pleasure to thank Jonathan Haidt, Jesse Graham and the 
YourMorals.org team
for letting us play with their questionnaires data, 
Vitor B P Leite for discussions about possible genetic basis of political ideology, 
Marcus V Baldo for discussions about learning and 
Paul Nail for comments on threats. 
This research received  support from the  Center for the Study of
 Natural and Artificial Information Processing Systems of the
University of S\~ao Paulo (CNAIPS NAP-USP). 
J Cesar received support from a FAPESP fellowship.

\newpage
\appendix

\section{The model and methods \label{model}}
A short description of the learning theory is presented below and
in the following appendices. 

Each agent is endowed with a learning system and
a set of weights. They exchange information, learning and teaching at
different instances, about a set of issues, represented each by a set
of numbers. Each number represents the bearing of the issue on
one of the moral dimensions. 
The dimension of the moral space, according to MFT
is around five or six. The development of the mathematical theory can be done 
for a general number of dimensions $d_m$. A choice has to be made
in order to compare with data, and since we are only comparing
with a set of questionnaires of conservatives and liberals, will use $d_m=5$
in the numerical part of the calculations. 
So, a statement or issue to be morally judged, at a time 
labelled by $\mu$ is represented as 
a vector in moral space 
$\bo{x}_\mu=(x_{1,\mu},x_{2,\mu},x_{3,\mu},x_{4,\mu},x_{5,\mu})$ 
with five components. For a particular agent, call it $i$,
the moral state of the agent,
called the moral matrix in MFT, is also a vector 
$\bo{\omega_i}=(\omega_{1,i},\omega_{2,i},\omega_{3,i},\omega_{4,i},
\omega_{5,i})$.

Moral judgements are taken to be intuitive, fast, not based on intricate
rules.  We suppose opinions to be constructed by 
 the  average of the components
of the issue, weighted by the values of the moral dimensions:
$h_{i,\mu}=\sum_{a=1}^{d_m}{\omega}_{a,i}{x}_{a,\mu}$ is the opinion of agent $i$
about issue $\bo{x}_\mu$. Furthermore we introduce
 the sign, for or against, of the opinion $\sigma_{i,\mu}=
\mathrm{sign}(h_{i,\mu})$ about the issue. 

We model social encounters
when agent $i$ receives information $y_\mu=(\sigma_{j,\mu},\bo{x_\mu})$  
emitted by the social partner $j$.
Since
the length of the vector $x_\mu$ does not alter the opinions $\sigma$,
we take all issues to be unit length.

Call $D_\mu=\{y_1,y_2,....y_\mu \}$ the set of all such pairs received 
until that time.

To take into account our limited access to information we have
to use a probabilistic framework. 
 Let $P(\bo{\omega}|D_\mu)$ describe our knowledge of the vector
of moral dimensions  $\bo{\omega}$ conditional on the information
the agent received until now $D_\mu$, composed of all the pairs 
up to time $\mu$. Now a new pair $y_{\mu+1}$ is received and 
the probability of having a particular moral dimension  $\bo{\omega}$ 
changes. That is the essence of learning. The basic relation of 
inference is drawn from Bayes theorem. If  $P(\bo{\omega}|D_\mu)$ is
the probability posterior to the consideration of the data set $D_\mu$
and prior  to the inclusion of the information contained
in the pair $y_{\mu+1}$, the basic assumption in 
 Bayesian learning is to use the old
posterior $P(\bo{\omega}|D_\mu)$  as the new prior. 
Then we can write for the updated distribution of the receiving agent
\begin{linenomath}
\begin{equation}
P(\bo{\omega}|D_{\mu+1})\propto P(\bo{\omega}|D_\mu)
P(\sigma_{i,\mu+1}=\sigma_{j,\mu+1}|\bo\omega,D_\mu, \bo{x_{\mu+1}})
\end{equation}
\end{linenomath}
The likelihood $P(\sigma_{i\,\mu+1}=\sigma_{j,\mu+1}|\bo\omega,D_\mu, \bo{x_{\mu+1}})$
describes the probability that  agent $i$
would have opinion $\sigma_{i\,\mu+1}=\sigma_{j,\mu+1}$  
about issue $\bo{x_{\mu+1}}$ if its
 moral vector were $\bo\omega$. 

For simplicity we consider an approximation where
the probability distributions are multivariate 
Gaussians. This family can be described by two objects:
a mean vector ($\bo{\hat{\omega}}$)  and a covariance matrix (${\bm C}$). Now
 the dynamics of learning can be simply
written by giving  the changes in these two 
quantities due to the incorporation of the information in the
example  $y_{\mu+1}$. After some manipulations (see \ref{app} below and \cite{Opper96}),
 the learning dynamics 
of agent $i$ is described 
 in terms of the components by
\begin{eqnarray}
\hat{\omega}_{a,\mu+1} &=& \hat{\omega}_{a,\mu} -
\sum_b C_{ab,\mu}\cdot \frac{\partial {\cal E}_\mu}{\partial \hat{\omega}_{b\mu}},
\label{dinamica1}\\ 
C_{ab,\mu+1} &=& C_{ab,\mu}-\sum_{cd} C_{ac,\mu} C_{bd,\mu}\frac{\partial^2
{\cal E}_\mu}{
\partial {\hat{\omega}_{c,\mu}}\partial {\hat{\omega}_{d,\mu}}}\label{dinamica2}
\end{eqnarray}
and
${\cal E}_\mu$, that can be called the learning energy or cost or pain, is given by
\begin{linenomath}
\begin{equation}
{\cal E}_\mu= -\ln \av {P(h_\mu| {\hat{\bo \omega}_\mu} +\bo u)}
\end{equation}
\end{linenomath}
where $h_\mu=\sum_a\hat{\omega}_{a,\mu}  x_{a,\mu+1}$  is the opinion
of the agent about issue $ x_{\mu+1}$  before receiving the opinion
of the social partner.
The average, represented by the angular brackets,
 is over the gaussian variable $\bo u$ with zero mean and 
covariance $C_{ab,\mu}$. Note that $\av {P(h_{\mu}|\hat{\bo\omega}+\bo u)}_\mu$
is also called the evidence. 
It is in the likelihood that enters the information about
how an issue and a moral vector give rise to an opinion and the noise
process that is corrupting the communication.

\subsection{Bayesian learning dynamics in the formative phase}
Different types of noise can enter in the communication
process. Here we suppose the case of multiplicative noise where
 a fraction $\epsilon$ of the opinions are inverted. 
The form of the learning potential
can be written as 
\begin{linenomath}
\begin{equation}
{\cal E}_{\mu}(z) =\log\left(
\epsilon + (1-2\epsilon)
\Phi\left( \frac{z}{\bm x^T_{\mu+1} \bm C_{\mu} \bm x_{\mu+1}}\right)\right)
\label{eq:ptau3}
\end{equation}
\end{linenomath}
where $z=\sigma_{j,\mu+1} h_{i,\mu}$ and 
$\Phi$ is the cumulative distribution of the gaussian ${\cal N}(0,1)$.
To simplify the interpretation of the results, at the expense
of small degradation in the performance of 
the learning algorithm we consider the case where the covariance has the
the from ${\bm C}_\mu= C_\mu{\bm 1}$, an overall factor $C_\mu$ 
times a unit matrix.
In this approximation  $\bm x^T_{\mu+1} \bm C_{\mu} \bm x_{\mu+1}=C_\mu$.
Then the dynamics is
\begin{eqnarray}
\hat{\omega}_{a,\mu+1} &=& \hat{\omega}_{a,\mu} - x_{a,\mu+1} \sigma_{j,\mu+1}
C_\mu \frac{\partial{\cal E}_\mu}{\partial z} , \\ 
C_{\mu+1} &=& C_\mu- C^{2}_\mu \frac{\partial^2{\cal E}_{\mu}}
{\partial z^2}. 
\end{eqnarray}

This dynamics and variations for other  learning scenarios
has been extensively analyzed in \cite{Kinouchi92,Opper96,Kinouchi93,Copelli95,Vicente98,biehl1995noisy,biehl1995gradient,caOl01,OlCa10,EngelBroeck}. 
We now make some comments that are relevant for our present
purposes. 

\subsection{The learning algorithm}
\begin{figure}[ht]
\begin{minipage}[b]{0.5\linewidth}
\centering
\includegraphics[scale=0.45]{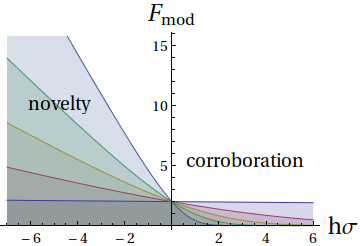}
\end{minipage}
\hspace{0.3cm}
\begin{minipage}[b]{0.5\linewidth}
\centering
\includegraphics[width=0.7\textwidth]{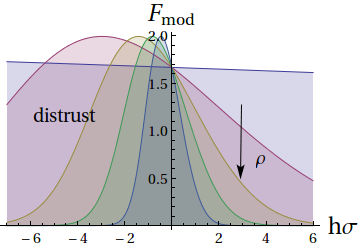}
\end{minipage}
\caption{\small {\bf The modulation function depends on the  a measure of novelty
and on the number of social opinion exchanges.} Left panel: Total confidence
on the information received. Right panel: Agent suspects that some
received opinions have been randomly inverted (20$\%$). 
The vertical axes are the modulation functions
and in the horizontal axes appears the product of the prior opinion
of the agent ($h$) times the sign ($\sigma$) 
of the arriving opinion information. 
Social interactions with opinions where 
 $h\sigma<0$ bring new information, those with
 $h\sigma>0$ are corroborative. 
The different curves are drawn each for different numbers of total
opinions to which the agent has been exposed, measured by $\rho$ which
increases as shown by the arrow ($\downarrow$).
The modulation function
changes from almost a constant, for very small number of social opinion
exchanges, to a very asymmetrical form where 
repetitive information causes
almost  no change at all and  novelty gives rise to 
a very high modulation of changes, except when a level of distrust has been
surpassed, as in the very negative region of $h\sigma$ in the right panel. 
See Appendix \ref{app} for details.  }
\label{fmod}
\end{figure}

We introduce  the modulation function (figure \ref{fmod})
$F_{mod}(z)=-C_\mu \frac{\partial{\cal E}_\mu}{\partial z} $ and write the 
dynamics as
\begin{eqnarray}
\hat{\omega}_{a,\mu+1} &=& \hat{\omega}_{a,\mu} +x_{a,\mu+1} \sigma_{j,\mu+1}F_{mod}(z_\mu),
\label{eqdinafmod}
\\
C_{\mu+1} &=& C_\mu+C_\mu \frac{\partial F_{mod}(z_\mu)}
{\partial z_\mu}. 
\end{eqnarray}
Learning is now seen as a modulated Hebbian learning,
where changes in the weights are done in the direction of
the vector $x_{\mu+1}$, if the social partner's opinion 
  $\sigma_{\mu+1}$ about it is positive 
and in the opposite direction it the opinion is negative. 
In Figure \ref{fmod} in the main text, the modulation function 
\begin{linenomath}
\begin{equation}
F_{mod}(z_{\mu}) =\frac
{(1-2\epsilon)
\exp-\frac{z_{\mu}^2}{2 C_{\mu}^2 }}
{\epsilon + (1-2\epsilon)
\Phi\left( \frac{z_{\mu}}{ C_{\mu} }\right)}
\label{modulationfunction}
\end{equation}
\end{linenomath}
is plotted as 
a function of $z$. Note that $z$ takes positive values if the
opinion of the agent and its social partner are the same
and is  negative if there is disagreement. If the absolute
value of $z$ is large the agent can be said to be very sure about
its opinion since small changes in the issue will not change its
classification. 

But more strikingly, the modulation function depends on $C$. 
In Figure \ref{fmod} we present $F_{mod}(z)$ for different values
of $\rho= 1/\sqrt{1+C^2}$, a convenient variable since it
takes values between zero and one.  It is close to 
zero when the agent's opinion has probability around  one half
of agreeing with that of the social partner. As learning occurs,
$\rho$ increases towards one. It can be shown that $\rho$ is related to
the probability $e_g$ of the opinions being different on a random issue, 
and $e_g$ goes to zero 
as $\rho \rightarrow 1$. In particular
$e_g=\frac{1}{\pi} \mathrm{accos}^{-1}\rho$ for large $d_m$ and uniform 
and independently distributed examples and it remains a useful variable 
in other conditions.

\subsection{Social influence phase}
We consider that the information exchanges
in the formative phase occur at  random and thus the effective $\rho$ for
each agent is a random number. Now we freeze the evolution 
of the modulation function, $\rho$ or equivalently  $C$ is fixed
at a particular value for each agent. We consider the agents 
to start a new phase in their lives where the value of
$\rho$ does not change anymore. The agents in the formative phase
learned to learn and now they just learn from each other. The validity
of this  supposition as something that
represents the developments of adolescents  has to be investigated
in an independent way. 

The dynamics of information exchange is analogous to that considered in 
\cite{Caticha2011, Vicente2012}, the only difference being that 
the learning occurs with the Bayesian algorithm described above. 

We suppose that a society discusses a set of $P$ issues. 
Parsing of an issue into a vector might be {\it subjective},
expressed by the fact that agent $i$ obtains a vector $\bm x_i$. 
Exchange of information between 
agents is about the average vector
\begin{linenomath}
\be
\z \propto \frac{1}{P}\sum_{\mu=1}^P \bm x_i^\mu,
\ee
\end{linenomath}
which we suppose reasonable to be independent of the agent, since
fluctuations due to subjective parsing, if unbiased, tend to cancel out. 
We call $\z$ the Zeitgeist vector since it captures the contributions
of all issues that are currently being discussed by the
society. Without any loss
it will be normalized to unit length. 
The opinion of agent $k$ about the
Zeitgeist is 
\begin{linenomath}
\be
h_k= \z \cdot \bm w_k
\ee
\end{linenomath}
and its  sign  is denoted by $\sigma_k=\mathrm{sign}(h_k)$. 
We now consider a Metropolis-like stochastic dynamics of information exchange.
Pick at random one agent, call $i$. Pick its social partner, call it $j$
uniformly  from its social neighbors.  
Now choose a $d_m$ dimensional vector  $\mathbf{u} $ drawn  uniformly 
on a ball of radius $\kappa$.   
A trial weight vector is defined by
\begin{eqnarray}
{\mathbf{T}}&=&\frac{\bm{w}_i(t) +\bm{u}}{|\bm{w}_i(t) +\bm{u}|} 
\end{eqnarray}
and accepted as the new weight vector, 
 $\bm{w}_i(t+1)=\bm{T}$ if the learning energy
: $\Delta {\cal E}:= {\cal E}(\mathbf{T},\sigma_j)- 
{\cal E}(\mathbf{w}_i(t),\sigma_j) \le 0$. If  $\Delta {\cal E} > 0 $
the change is accepted with probability $ \exp -\beta\Delta {\cal E} $.
In analogy to equation  \ref{eq:ptau3}
\begin{linenomath}
\begin{equation}
{\cal E}(\bm w_i,\sigma_j) =\log\left(
\epsilon + (1-2\epsilon)
\Phi\left( \frac{h_i \sigma_j}{C}\right)\right)
\label{metropolis}
\end{equation}
\end{linenomath}
This is looped randomly over the whole population.  
A technical comment is that it is not obvious if
this dynamics leads to an equilibrium state since the
energy ${\cal E}$ is not symmetric with respect to the interchange
of the two actor agents: the emitter and the receiver of the information.
However numerically, 
order parameters rapidly converge to values that remain stationary during
thousands of iterations, a time scale that we consider sufficient to 
study a steady state and further to consider the effects of Zeitgeist
changes and readaptations. Within certain limits $\kappa$ controls
the acceptance rate and thus the time scales to reach stationary values.

The value of $\beta$ sets the scale of fluctuations of the 
energy ${\cal E}$. If $\beta$ is large, even  small changes 
$\Delta {\cal E}$ have large effects and large changes will 
not be possible, and if $\beta$ is small,
then large fluctuations may be easily accepted. Our interpretation 
is that $\beta$ serves as a pressure to accommodate and conform to the
opinion of others. 
Large $\beta$ means strict conformity, while in a small $\beta$ 
regime, tolerance to fluctuations in conformity are accepted.

\section{Bayesian inspired learning algorithms \label{app}}
For the learning set  $D_t = \left( y_0,\ldots,y_{t-1} \right)$ 
of independently chosen vectors and their opinions, the likelihood
is a product 
\begin{linenomath}
\be
    P(D_t|\bm{\omega}) = \prod_{i=0}^{t-1} P(y_i|{\bm\omega})
\ee
\end{linenomath}
where  ${\bm\omega}$ is the set parameters to be inferred. The data comes in ordered
pairs
$y_t = ( \sigma_{j,t}, \bm x_t )$ where $\sigma_{j,t}$ is positive if
agent $j$ considers issue $\bm x_t$ as a morally acceptable issue and negative
otherwise; 
$ \bm x_t = (x_t^1, \ldots,
x_t^N) $ is a five dimensional vector. Our choice of $N=5$ is determined by 
Moral Foundation theory. 
 
Bayesian inference derives from the application of Bayes theorem in order to 
incorporate information that permits updating from a {\it prior} to a 
{\it posterior} distribution:
\begin{linenomath}
\be
    P({\bm \omega}|D_t) = \frac{P({\bm \omega})P(D_t|\bm \omega)}
        {\int \dd {\bm\omega}' P({\bm \omega}')P(D_t|{\bm \omega}')}
\ee
\end{linenomath}

In Online learning we consider the updating of the distribution
due to the addition of a single example pair 
$y_{t + 1}$ 
\begin{linenomath}
\be
    P({\bm\omega}|D_{t+1}) = 
        \frac{P({\bm\omega}|D_t)P(y_t|{\bm\omega})}
        {\int d {\bm\omega}' P(\bm{\omega}'|D_t)P(y_t|\bm{\omega}')}.
    \label{eq:td1}
\ee
\end{linenomath}
The amount of memory needed to store the whole posterior can be prohibitively
large and following Opper \cite{Opper96} we consider a simplification where the
posterior is constrained to belong to a parametric family, which we take
to be the $N$ dimensional multivariate Gaussian.

If at a certain stage our knowledge is codified into one such Gaussian,
\begin{eqnarray}
    P_G(\bm\omega|D_t) &= &
        \frac{1}{|2\pi \det C_t|^{\half}}
        \exp \left( - \frac{1}{2} \left(\bm \omega - \hat{\bm \omega}_t 
\right)^T
         \bm C_t^{-1}\left(\bm \omega - \hat{\bm \omega}_t \right)\right)
\end{eqnarray}
a Bayesian update will in general take the posterior
 out of the Gaussian space. Then a new Gaussian posterior is chosen 
is such a way that the information loss is minimized. Thus the learning
step is comprised of two sub-steps:
\begin{itemize}
    \item \textbf{New example drives the posterior out of the Gaussian space }:
    \begin{linenomath}
    \begin{equation}
      P(\bm \omega|D_{t+1}) :=    P(\bm \omega|D_t , y_t) = 
            \frac{P_G(\bm\omega|D_t)P(y_t|\bm\omega)}
            {\int \dd \omega' 
            P_G(\bm\omega'|D_t)P(y_t|\bm\omega')}
        \label{eq:td2}
    \end{equation}
    \end{linenomath}
    \item \textbf{Project back to Gaussian space}: 
    \begin{linenomath}    
    \begin{equation}
        P(\bm\omega|D_{t+1}) 
        \rightarrow
        P_G(\bm\omega|D_{t + 1}) 
    \end{equation}
    \end{linenomath}
The projection step is done using the Kullback-Leibler divergence or
equivalently, by maximizing the cross entropy:
    \begin{linenomath}
    \begin{eqnarray}
         & KL[ P(\bm\omega|D_{t+1})||P_G(\bm\omega|D_{t + 1}) ] \nn
         &\qquad =  \int \dd \omega P(\bm\omega|D_t , y_t)
         \log \frac{P(\bm\omega|D_{t+1})}{P_G(\bm\omega|D_{t + 1})}
    \label{eq:kl1}
    \end{eqnarray}
    \end{linenomath}
\end{itemize}
The minimization of the KL divergence results in projecting into 
 the Gaussian with the same mean and covariance vector as the non-Gaussian posterior:
\begin{eqnarray}
     \hat{\bm \omega}_{t+1} &= \frac{\int \dd \bm \omega \bm \omega P(\bm \omega|D_{t+1})P(y_t|\bm \omega)} 
                          {\int \dd \bm \omega P(\bm \omega|D_{t+1})P(y_t|\bm \omega)} \nn
     \bm C_{t+1}     &= \frac{\int \dd \bm \omega \bm \omega \bm \omega^T P(\bm \omega|D_{t+1})P(y_t|\bm \omega)}
                          {\int \dd \bm \omega P(\bm \omega|D_{t+1})P(y_t|\bm \omega)}
                    -\hat{\bm \omega}_{t+1}  \hat{\bm  \omega}_{t+1}^T
     \label{eq:iter1}.
\end{eqnarray}
Now change variables, introducing $\bm u$ the fluctuations around the mean
$\bm \omega =  \hat{ \bm \omega}_t+ \bm u$. Using that for Gaussians
with zero mean
$\E(xf(x)) = \E(x^2)\E(f'(x))$
and $ \td{f(x + y)}{x} = \td{f(x + y)}{y}$ then it follows that
the new mean and covariance change as described by equations \ref{dinamicaE},
\ref{dinamicaF} (main text), \ref{dinamica1} and \ref{dinamica2} in \ref{model}.

\section{Comparing ERN and the  modulation function \label{ERNrho}}
\begin{figure}[ht]
\centering
\includegraphics[scale=0.6]{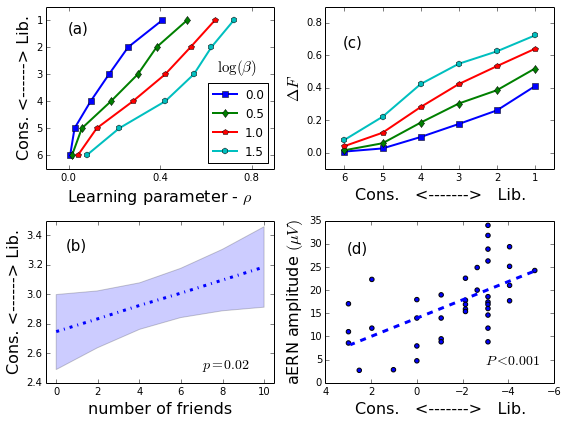}
\caption{{\bf Model versus experiment.} Model results: panels (a) and (c).
Experimental results from data published  in  \cite{Settle2010b} 
(panel (b)) and in \cite{Amodio2007} (panel (d)). 
(a) Model: Political affiliation (pa= 1-Liberal, pa=7-Very Conservative)
is correlated to $\rho$, which is a measure of the number of social 
information exchanges in the formative phase of the agents' lives. This occurs
for a wide range of pressures $\beta$ values.  
(b) Data: Number of friendships in people with the two alleles of DRD4-7R 
correlates with liberalism \cite{Settle2010b}.
(c) Model: Difference between average modulation in novel and corroborative 
situations, 
 $\Delta F=\av {F_{mod}}_{novelty} - \av{F_{mod}}_{corroboration}$.
 correlates with liberalism (see section \ref{ERNrho}.)
(d) Data: magnitude of ERN difference between novelty and corroboration in 
go-no-go game from   \cite{Amodio2007} correlates with liberalism.
}
\label{politicalaffiliation}
\end{figure}

The modulation function determines the size of the weight changes
during learning. We define the average of the modulation function 
for novelty $\av {F_{mod}}_{novelty}$ and  $\av {F_{mod}}_{corroboration}$
by
\begin{eqnarray}
\av {F_{mod}}_{novelty}&=&\av F ^{-1} \int_{-\infty}^0F_{mod}(z)P(z)dz, \nonumber\\
\av {F_{mod}}_{corroboration}&=&\av F ^{-1} \int_0^{\infty}F_{mod}(z)P(z)dz,\nonumber\\
\av F&=&\int_{-\infty}^\infty F_{mod}(z)P(z)dz\nonumber.
\end{eqnarray}
For a unifomr distribution of examples, and 
with the normalization of $\bm\omega$,  the
distribution $P(z)$ of $z=h\sigma$, 
is  the gaussian distribution with zero mean and unit variance.
Since the modulation function depends on $\rho$, the difference
\begin{linenomath}
\begin{equation}
\Delta F=\av {F_{mod}}_{novelty}-\av {F_{mod}}_{corroboration}
\end{equation}
\end{linenomath}
can be identified to a political affiliation. This is shown 
in figure \ref{politicalaffiliation}.c.
This is the closest we can come theoretically to defining 
within the model a quantity similar to the Error Related Negativity
(ERN) 
measured by Amodio {\it et al} \cite{Amodio2007} which reports 
differences between measured EEG signals of unexpected
and expected situations conditional on self-declared
political affiliations. In figure \ref{politicalaffiliation}.d
we show the results from \cite{Amodio2007} for
 the magnitude of the ERN signal versus political affiliations.

\bibliography{cacevibib}


\end{document}